\pdfoutput=1

\documentclass[twocolumn,showpacs,amssymb,aps,nofootinbib,floatfix,superscriptaddress]{revtex4-1}

\usepackage{epsfig}
\usepackage{hyperref}
\usepackage{comment}

\usepackage{graphicx,color}
\begin{document} 
\hbadness=10000

\title{Hydrodynamic modeling of pseudorapidity flow correlations in relativistic heavy-ion collisions and the torque effect}

\author{Piotr Bo\.zek}
\email{Piotr.Bozek@fis.agh.edu.pl}
\affiliation{AGH University of Science and Technology, Faculty of Physics and
Applied Computer Science, al. Mickiewicza 30, 30-059 Krakow, Poland}

\author{Wojciech Broniowski}
\email{Wojciech.Broniowski@ifj.edu.pl}
\affiliation{The H. Niewodnicza\'nski Institute of Nuclear Physics, Polish Academy of Sciences, 31-342 Cracow, Poland}
\affiliation{Institute of Physics, Jan Kochanowski University, 25-406 Kielce, Poland}

\author{Adam Olszewski}
\email{Adam.Olszewski.fiz@gmail.com}
\affiliation{Institute of Physics, Jan Kochanowski University, 25-406 Kielce, Poland}

\begin{abstract}
We analyze correlations between the event-plane angles in different intervals of pseudorapidity within the 3+1-dimensional 
viscous hydrodynamics with the Glauber-model initial conditions. 
As predicted earlier, the fluctuations in the particle production mechanism in the earliest stage, together with asymmetry of the 
emission profiles in pseudorapidity from the forward- and backward going wounded nucleons, lead to the {\em torque} 
effect, namely, decorrelation of the event-plane angles in distant pseudorapidity bins.
We use 
%two methods to identify decorrelation of the event
%planes, the principal component analysis and the 
two- or three-bin measures of correlation functions to quantify the effect, 
with the latter compared to the recent data from the CMS collaboration.
%The subleading modes in 
%the principal component analysis are   stron
%rapidity intervals is due to both  the torque effect and to non-flow effects. 
We find a sizable torque 
%for the third order event plane for all considered centralities, and for the second order event plane in central
%collisions. The 
effect, with magnitude larger at RHIC than at the LHC.
\end{abstract}



\maketitle

%%%%%%%%%%%%%%%%%%%%%%%%%%%%%%%

\section{Introduction \label{sec:intro}}

Fluctuations of the shape of the fireball created in relativistic heavy-ion
collisions give rise to event-by-event fluctuations 
of the collective flow.  The main examples of such phenomena are 
the elliptic flow fluctuations~\cite{Alver:2007rm}, 
the triangular flow~\cite{Alver:2010gr}, the mean transverse momentum 
fluctuations~\cite{Broniowski:2009fm}, the broad event-by-event distribution
 of the flow coefficients~\cite{Aad:2013xma}, or breaking of the factorization of the 
 two-particle correlations~\cite{Gardim:2012im}. 
 The fluctuations 
 %in the spectra of the emitted particles can
 %be described in the hydrodynamic model, shape fluctuation 
 of the initial distribution of sources give rise, due 
 to the hydrodynamic expansion, to event-by-event fluctuations of the 
 collective flow~\cite{Schenke:2010rr,*Bozek:2012fw,*Gale:2012rq,*Heinz:2013bua}.

A related aspect is the rapidity dependence of the collective flow. Two-particle correlations in relative 
pseudorapidity and relative azimuthal angle exhibit ridge-like structures extended in the longitudinal direction. 
%\cite{Agakishiev:2011pe,Chatrchyan:2011eka,...}. 
These long-range fluctuations in rapidity indicate the presence of longitudinal
correlations in the initial stage of the collision. The approximate preservation of the transverse shape of the fireball 
along the longitudinal direction yields strong correlations of the
flow event-plane angles at different rapidities, which allows one to extract the flow coefficients using 
the event plane direction defined in a forward rapidity interval.

A mechanism leading to decorrelation of event-planes at different rapidities, the {\em torque} effect, has been proposed in Ref.~\cite{Bozek:2010vz}.
One expects that at forward (backward) rapidities the  event-plane direction is predominantly determined by the participant nucleons
from the forward (backward) going wounded nucleus. While strongly correlated, the event-plane directions defined by the forward (backward) 
going wounded nucleons are different due to event-by-event fluctuations. At intermediate rapidities the shape of 
the interaction region should interpolate between the  forward and backward limits. As a result, the orientation of the
event-plane changes along pseudorapidity in a given event: the torque bends to the left or right, with a fluctuating angle. 
The details and the strength of the torque effect depend on the assumed mechanism for the initial energy 
deposition in the transverse plane and in space-time rapidity. Thus, the correlations in the longitudinal direction can serve as a tool to study the 
mechanisms of the initial energy deposition in the fireball. 

In our original study~\cite{Bozek:2010vz}, concerning the Au+Au collisions at 
the Brookhaven Relativistic Heavy-Ion Collider (RHIC), 
we have not carried out the event-by-event analysis, which is presented here. Instead, we have 
used the fact that the standard deviation of the torque angle between the forward and backward directions in the Glauber initial condition has a particular value (of the order of 10$^o$ for the 
pseudorapidity separation of 2 units). This value was implemented in the initial shape, later evolved with hydrodynamics. We have also carried out the statistical hadronization at freeze-out 
with {\tt THERMINATOR}~\cite{Kisiel:2005hn,*Chojnacki:2011hb} to generate the distribution of the observed hadrons. The finite number of the produced hadrons leads to purely 
statistical fluctuations of the event-plane angles, which must be carefully sorted out through appropriate torque-effect measures.
Ref.~\cite{Bozek:2010vz} also provided the estimates for two- and four-particle measures of the effect, as well as explored the 
centrality dependence of the effect at the level of the Glauber initial state.

In this work a full-fledged hydrodynamic event-by-event analysis of the torque effect is presented and compared to the results of the CMS 
Collaboration~\cite{Khachatryan:2015oea}.
%We  show  results for a number of observables related to the longitudinal event-plane correlations
%for a specific model of initial conditions. 
Specifically, to prepare initial conditions we use the Glauber Monte Carlo model 
with the assumption of asymmetric emission in the space-time rapidity from the
target and projectile participants~\cite{Bozek:2010bi}, motivated with the parameterizations from 
Refs.~\cite{Bialas:2004su,*Gazdzicki:2005rr,*Bzdak:2009dr,*Bzdak:2009xq}. As mentioned above and discussed in detail 
in Ref.~\cite{Bozek:2010vz}, the asymmetry of the energy deposition in pseudorapidity gives the torque 
of the event-plane as a function of the space-time rapidity. A similar mechanism yields the finite correlation range of particle multiplicities
in pseudorapidity intervals as a function of the separation between the  intervals~\cite{Bzdak:2009xq}. 
We  note that multiplicity correlation in different pseudorapidity intervals have been the subject numerous experimental 
and theoretical investigations~%
\cite{Back:2006id,*Abelev:2009ag,*Feofilov:2013kna,*De:2013bta,*Dusling:2009ni,*Armesto:2006bv,*Fukushima:2008ya,%
*Amelin:1994mf,*Braun:1997ch,*Brogueira:2006yk,*Yan:2010et,*Bialas:2011xk,*Olszewski:2013qwa,*Bzdak:2012tp,*Olszewski:2015xba}.

The event-plane decorrelation as a function of the pseudorapidity separation has been identified in several models of the initial
conditions in relativistic heavy-ion collisions~\cite{Petersen:2011fp,*Xiao:2012uw,*Jia:2014ysa,*Jia:2014vja,*Pang:2014pxa}. First  
experimental indications  for the event-plane dependence on rapidity came from
the azimuthally sensitive Hanbury--Brown-Twiss analysis~\cite{Niida:2015ifa}. Very recently, the CMS Collaboration
measured the effect with the help of three-bin correlations~\cite{Khachatryan:2015oea}, which greatly reduces the sensitivity to non-flow effects.

%The basic difficulty resides in identifying
%the correct observable  for the average event-plane rotation angle, 
%insensitive to finite multiplicity fluctuations and non-flow correlations.

\section{Formalism \label{sec:form}}

We perform   event-by-event simulations for Au+Au collisions at
$\sqrt{s_{NN}}=200$~GeV (RHIC) and Pb+Pb collisions at $2.76$~TeV carried out at the Large Hadron Collider (LHC). 
To pinpoint the torque effect, we compare rapidity correlation for two sets of initial conditions: 
a forward-backward symmetric longitudinal profile of the deposited entropy for each wounded nucleon (no-torque, which serves as the 
reference calculation), and the profile with asymmetric deposition from the left- and right-going 
nucleons (torqued initial conditions). 
%
%We calculate correlation of event-planes in intervals of pseudorapidity as a function of the separation between the intervals. The comparison of the results for the two
%initial condition allows one to compare the decorrelation due to the twist in the event-plane angle imposed in the initial condition 
%to  taht from non-flow effects. For Au-Au and Pb-Pb collisions we perform the principal component analysis (PCA)
%decomposition of the multiplicity correlations and event-plane correlations. 
%
The subsequent expansion of the fireball is modeled with 3+1-dimensional viscous hydrodynamics. Hydrodynamic simulations are carried out
on event-by-event basis for every generated initial condition. The viscosity to entropy density ratio 
is $\eta/s=0.08$ for the shear viscosity (independent of temperature) and $\zeta/s=0.04$ for the bulk viscosity 
(for $T<170$~MeV)~\cite{Bozek:2009dw}. The fluid expands from the initial time $\tau_0=0.6$~fm/c until freeze-out at 
 the temperature $T_{F}=150$~MeV. At freeze-out, hadrons are emitted statistically and afterwards resonance decays occur, 
as implemented in the event generator {\tt THERMINATOR}~\cite{Chojnacki:2011hb}. For each investigated centrality we generate from 200 to 300 hydro events. 
To increase statistics,  from  200 to 2000 (depending on the energy and collision centrality) {\tt THERMINATOR} events are generated for each freeze-out hypersurface.

The initial entropy density in the fireball is calculated with {\tt GLISSANDO}~\cite{Broniowski:2007nz,Rybczynski:2013yba}. 
The positions $x_i, y_i$ of  the  $N_{+}$ ($N_-$) wounded nucleons~\cite{Bialas:1976ed} from  the forward- (backward-)going nucleus are generated in each event. 
We use a Gaussian wounding profile for the
nucleon-nucleon collisions in the Glauber model, and an excluded distance $d=0.9$~fm for the nucleon configurations
in the nuclei. The total inelastic nucleon-nucleon cross section is taken to be 42~mb at RHIC and 63~mb at the LHC.
The entropy density in the transverse plane and the space-time rapidity  $\eta_\parallel$ is modeled as
a sum of contributions from the individual nucleons
\begin{equation}
s(x,y,\eta_\parallel)=\sum_{i=1}^{N_+} g_i(x,y) f_+(\eta_+) + \sum_{i=1}^{N_-} g_i(x,y) f_-(\eta_\parallel)   \ .
\end{equation} 
The density in the transverse plane has a Gaussian form
\begin{eqnarray}
g_i(x,y)&=& \kappa \left[ (1-\alpha) + N^{\rm coll}_i \alpha \right] \times \nonumber \\
&& \exp\left( - \frac{(x-x_i)^2+(y-y_i)^2}{2\sigma^2}\right) ,
\end{eqnarray}
where the width of the Gaussian is $\sigma=0.4$~fm, $N^{\rm coll}_i$ is the number of collisions for the $i$-th nucleon,
and $\alpha=0.12$ and $0.15$ and RHIC and the LHC, respectively. 
We note that  $\sum_{i=1}^{N_+} N_i^{\rm coll} = \sum_{i=1}^{N_-} N_i^{\rm coll} =N_{\rm bin}$ -- the number of binary collisions.
The non-zero value of $\alpha$ implements the mixed  
Glauber model~\cite{Kharzeev:2000ph,*Back:2001xy}, adding binary collisions to the wounded nucleon model.
The parameter $\kappa$ is adjusted for 
each energy to reproduce the average particle multiplicity in central events.  

The longitudinal density profile (an approximate ``triangle'')
\begin{equation}
f_{\pm}(\eta_\parallel)= \frac{\eta_{\rm beam}\pm \eta_\parallel}{2 y_{\rm beam}} H(\eta_\parallel)\  \mbox {for } \ |\eta_\parallel|<y_{\rm beam} 
\label{eq:lprof}
\end{equation}
is a product of an asymmetric linear function and an overall rapidity profile $H(\eta_\parallel)$, taken 
in the form of a plateau with Gaussian tails~\cite{Hirano:2002ds},
\begin{equation}
H(\eta_\parallel)=\exp\left(-\frac{(|\eta_\parallel|-\eta_p)^2\Theta(|\eta_\parallel|-\eta_p)}{2\sigma_\eta^2}\right) ,
\end{equation}
where $\sigma_\eta=1.4$ and $\eta_p=0.75$ and $1.15$ at RHIC and the LHC, respectively.
The ansatz (\ref{eq:lprof}) 
with an asymmetric deposition of entropy is based on the analysis of asymmetric d-Au collisions at RHIC~\cite{Bialas:2004su}. 
The event planes defined with the forward- and backward-going wounded nucleons lead to different event-plane directions, due to fluctuations
of $N_+$ and $N_-$. The asymmetric deposition of  entropy makes the event plane direction rapidity dependent \cite{Bozek:2010vz}.

\begin{figure*}[tb]
\begin{center}
\includegraphics[width=0.35 \textwidth]{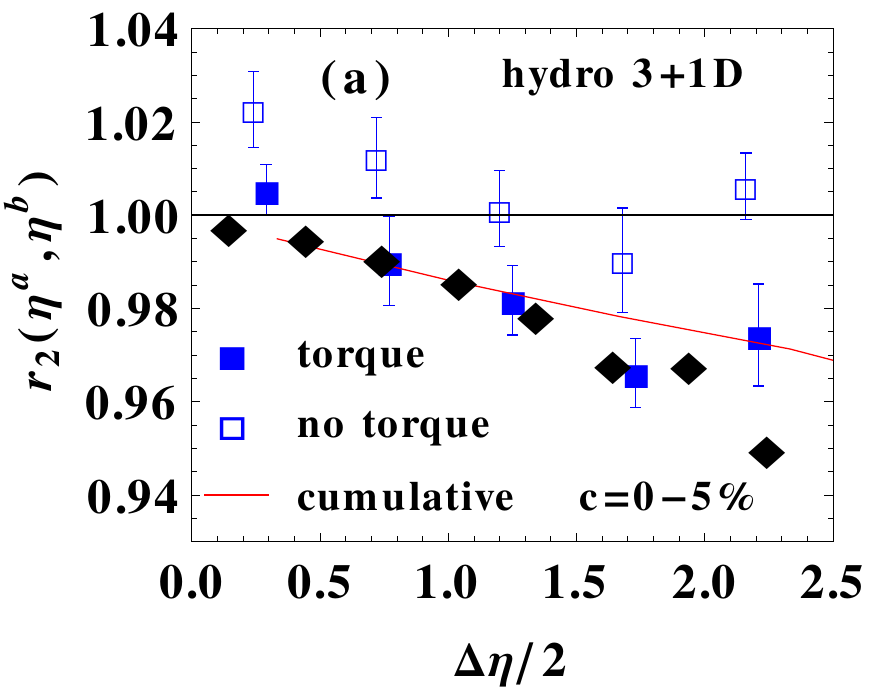} \hspace{7mm}
\includegraphics[width=0.35 \textwidth]{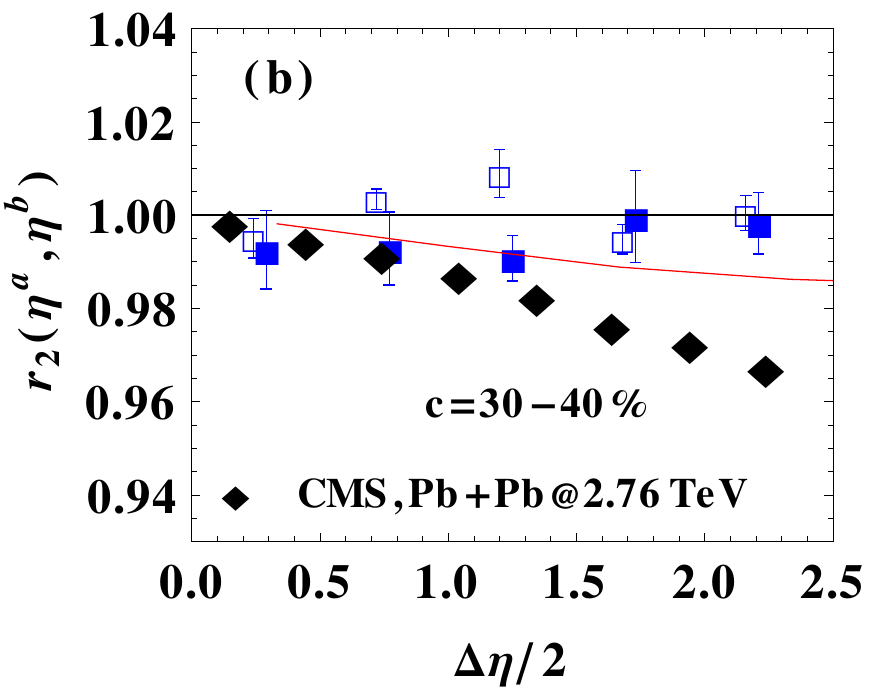}\\ \vspace{-6mm}
\includegraphics[width=0.35 \textwidth]{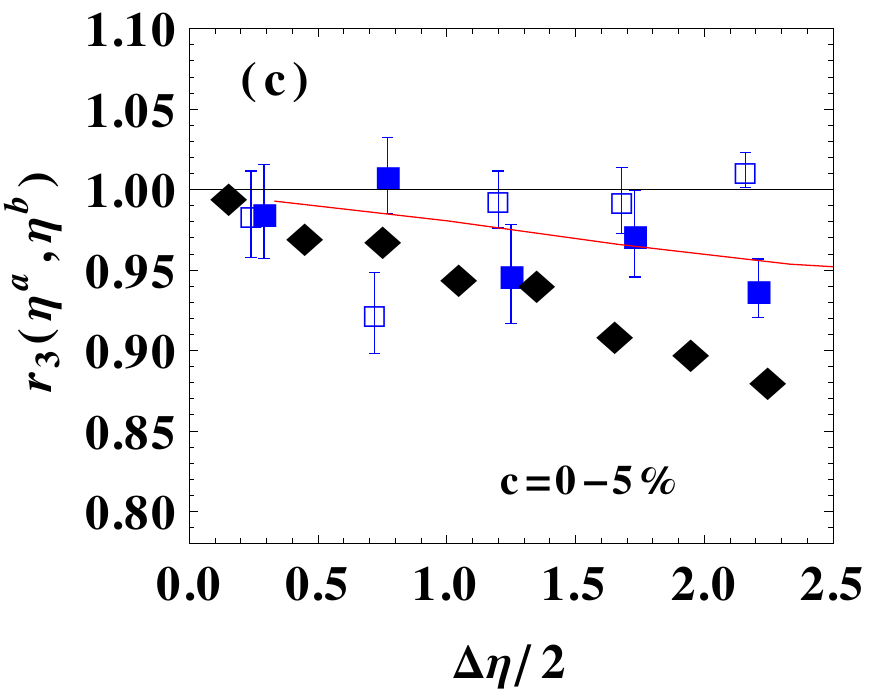} \hspace{7mm}
\includegraphics[width=0.35 \textwidth]{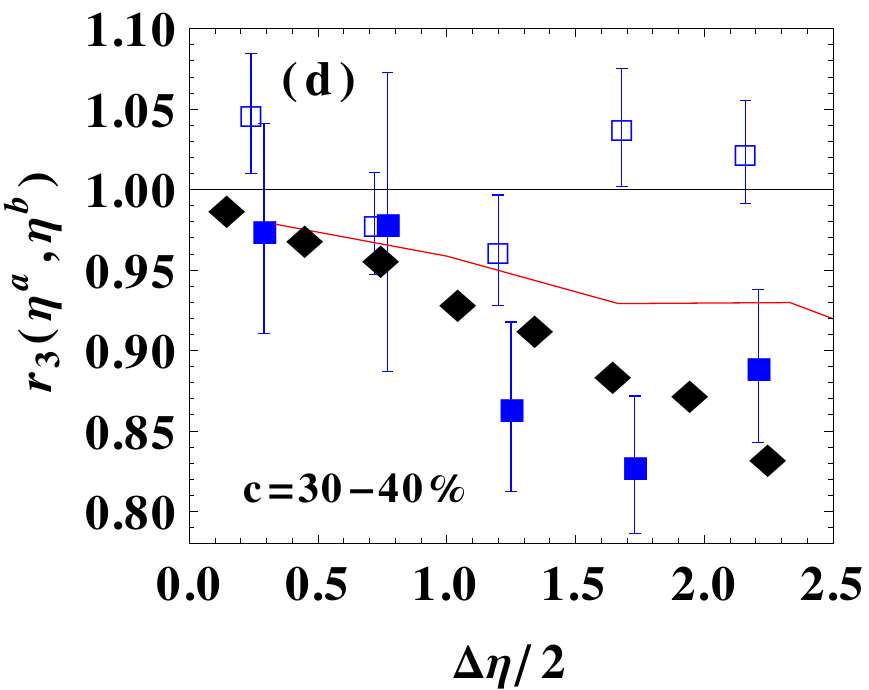}
\caption{(color online) The 3-bin measure for the elliptic flow of charged hadrons for two centralities and for elliptic ($n=2$) and 
triangular ($n=3$) flow. The CMS data come from Ref.~\cite{Khachatryan:2015oea}. The empty squares correspond to 
the model without the torque effect and are compatible with 1. The filled squares are the result of simulations with the torque effect. The solid line 
shows the calculation with cumulative events (see text for details). The model points are slightly displaced horizontally for better visibility.
\label{fig:3bin}}
\end{center}
\end{figure*}

We recall that the above formulas also lead to a {\em tilt} of the fireball away from the collisions axis (rotation in the $z-x$ plane), 
which generates the rapidity-odd directed flow~\cite{Bozek:2010bi}. %The asymmetric deposition of entropy introduces 
%a scale in the forward-backward multiplicity correlation  \cite{Bzdak:2009xq}.
The exact form of the asymmetric ansatz for the initial fireball can be varied, leading to slightly different 
predictions for the directed flow~\cite{Bozek:2010bi,Becattini:2015ska}. 
In the following we use the ansatz (\ref{eq:lprof}), which was shown to reproduce the directed 
flow at RHIC with the viscous hydrodynamic evolution using $\eta/s=0.08$~\cite{Bozek:2011ua}.

For comparison, we also perform simulations with a symmetric (no-torque, no-tilt) emission ansatz, where 
$f_{\pm}(\eta_\parallel)=H(\eta_\parallel)$.
All the simulation points in this paper are drawn with error bars estimated from resampling of the Monte Carlo events obtained
from {\tt THERMINATOR}.

\section{3-bin correlations \label{sec:3bin}}

We first show our results for the torque effect based on the three-bin correlation measure 
introduced by the CMS Collaboration~\cite{Khachatryan:2015oea}. It is defined as
\begin{eqnarray}
 r_n(\eta^a, \eta^b)=\frac{V_{n\Delta}(-\eta^a,\eta^b)}{V_{n\Delta}(\eta^a,\eta^b)}, \label{eq:rn}
\end{eqnarray}
where 
\begin{eqnarray}
V_{n\Delta}(\eta^a, \eta^b)=\langle \langle \cos[n(\phi_a-\phi_b)] \rangle \rangle.
\end{eqnarray}
The averaging is over all charged hadron pairs where one particle belongs to bin $a$ and the other to bin $b$, and over events.
The geometric interpretation of the above measure is very simple when the magnitude of the flow coefficient, $v_n(\eta^a)$,  is uncorrelated to the 
event-plane angle in bin $a$, $\Psi_n(\eta^a)$. In that case 
\begin{eqnarray}
 r_n(\eta^a, \eta^b)=\frac{\cos[n(\Psi_n(-\eta^a)-\Psi_n(\eta^b))]}{\cos[n(\Psi_n(\eta^a)-\Psi_n(\eta^b))]},
\end{eqnarray}

\begin{figure*}[tb]
\begin{center}
\includegraphics[width=0.35 \textwidth]{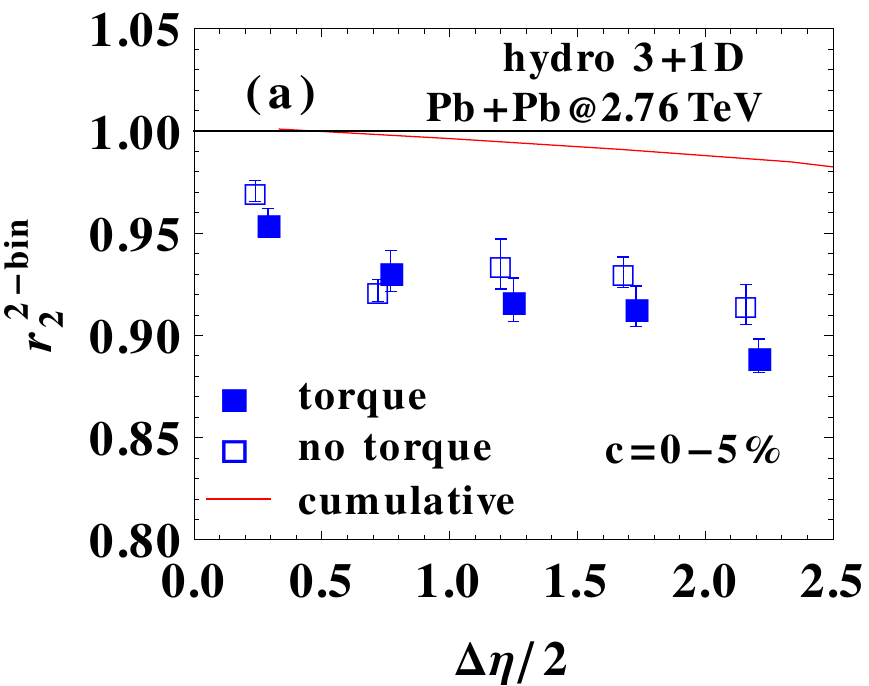} \hspace{7mm}
\includegraphics[width=0.35 \textwidth]{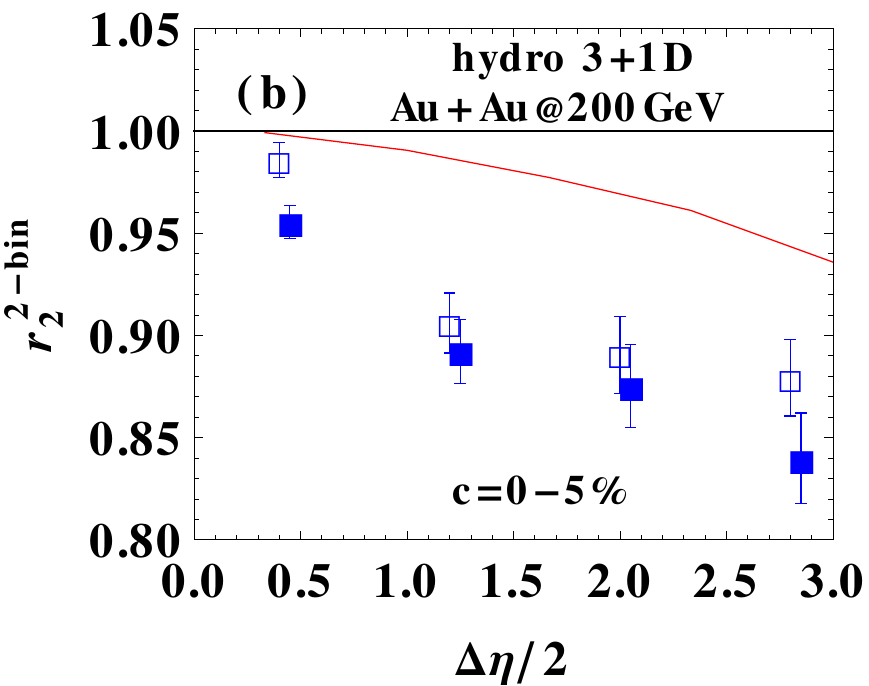}\\ \vspace{-6.5mm}
\includegraphics[width=0.35 \textwidth]{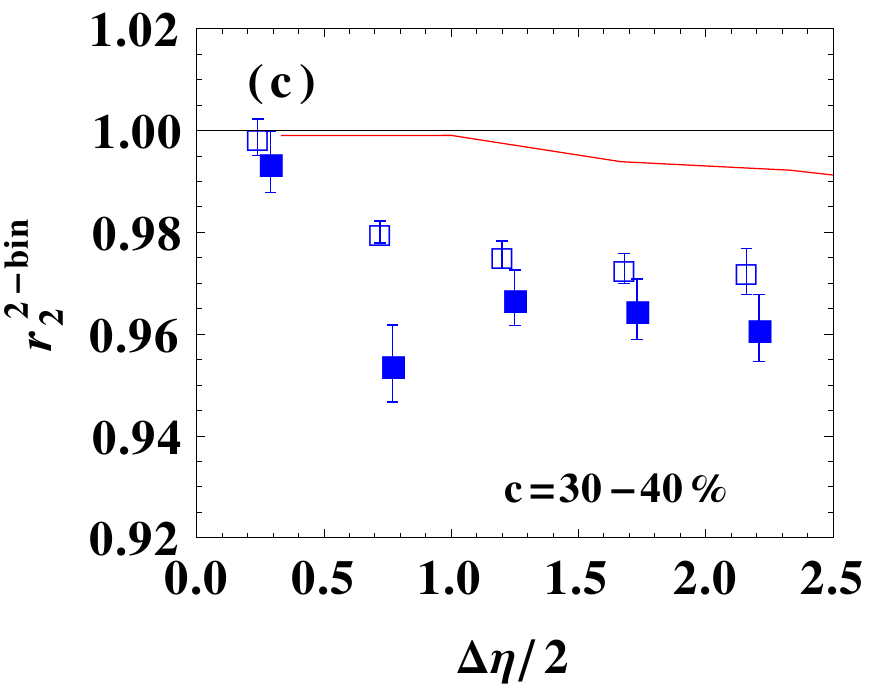} \hspace{7mm}
\includegraphics[width=0.35 \textwidth]{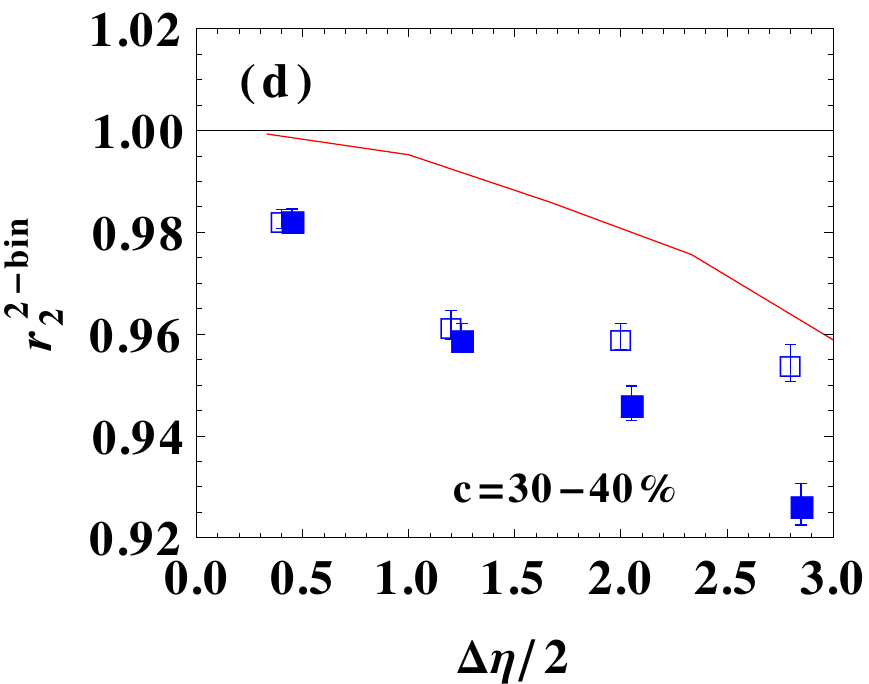}
\caption{(color online) Model predictions for the torque measure $r_2^{\rm 2-bin}$ for charged particles at 
the LHC (panels a and c) and RHIC (panels b and d) energies and for two sample centralities.  The empty squares correspond to 
the model without the torque effect. The filled squares are the result of simulations with the torque effect. The solid line 
shows the calculation with cumulative events.
\label{fig:tv2}}
\end{center}
\end{figure*}

An important feature of measure (\ref{eq:rn}) is that it cancels the random fluctuations of the event-plane angles, which result from 
a finite number of hadrons in the pseudorapidity bins. For distant bins $a$ and $b$, the fluctuation of  $(\Psi_n(\eta^a)-\Psi_n(\eta^b)$ is 
on the average the same as for $(\Psi_n(-\eta^a)-\Psi_n(\eta^b)$. Then, in the absence of a genuine torque effect, we would have 
$r_n(\eta^a,\eta^b)=1$. The departure of the measure from unity quantifies the torque effect.

\begin{figure*}[tb]
\begin{center}
\includegraphics[width=0.35 \textwidth]{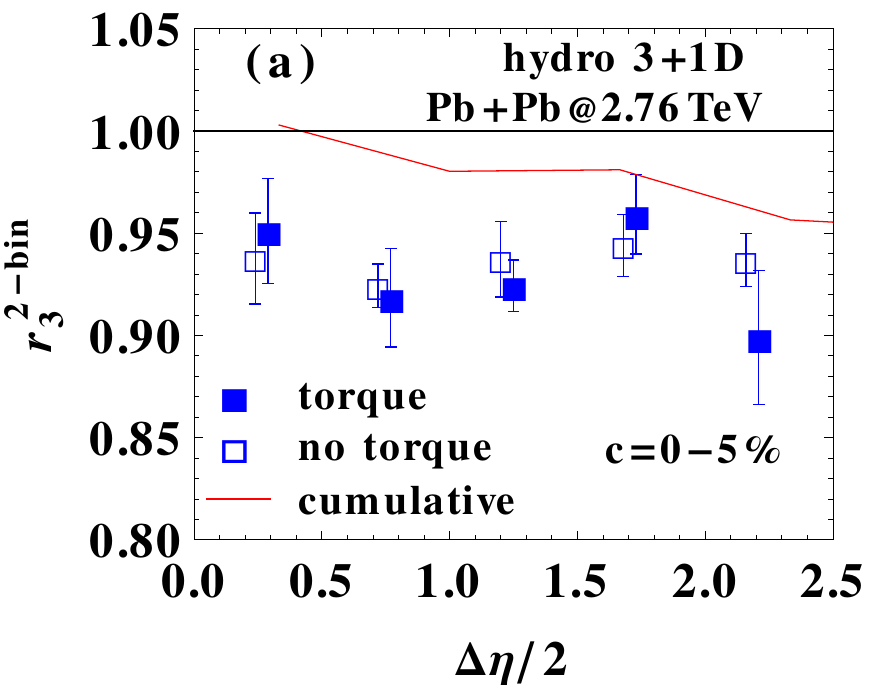} \hspace{7mm}
\includegraphics[width=0.35 \textwidth]{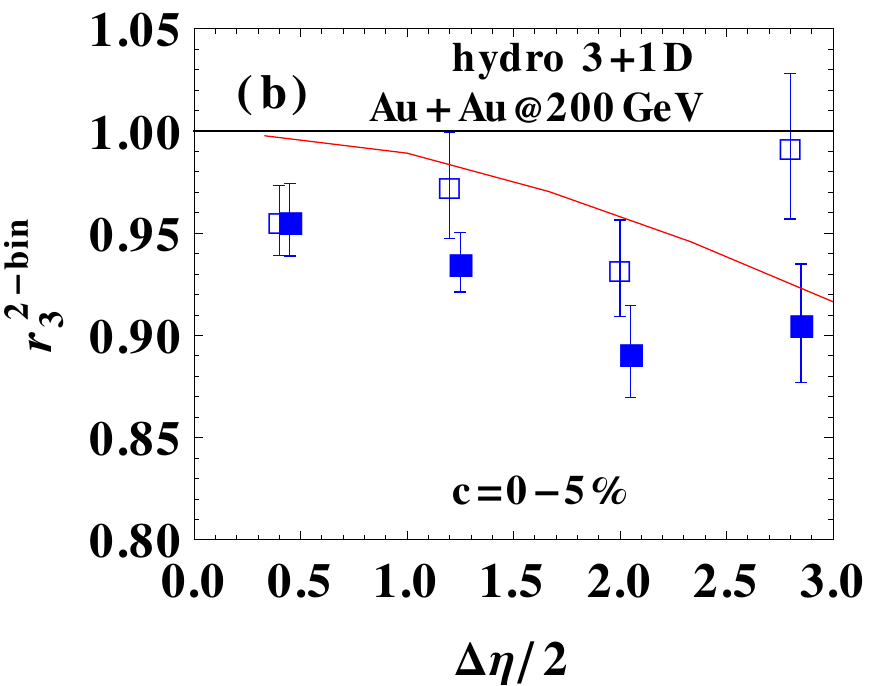}\\ \vspace{-6.5mm}
\includegraphics[width=0.35 \textwidth]{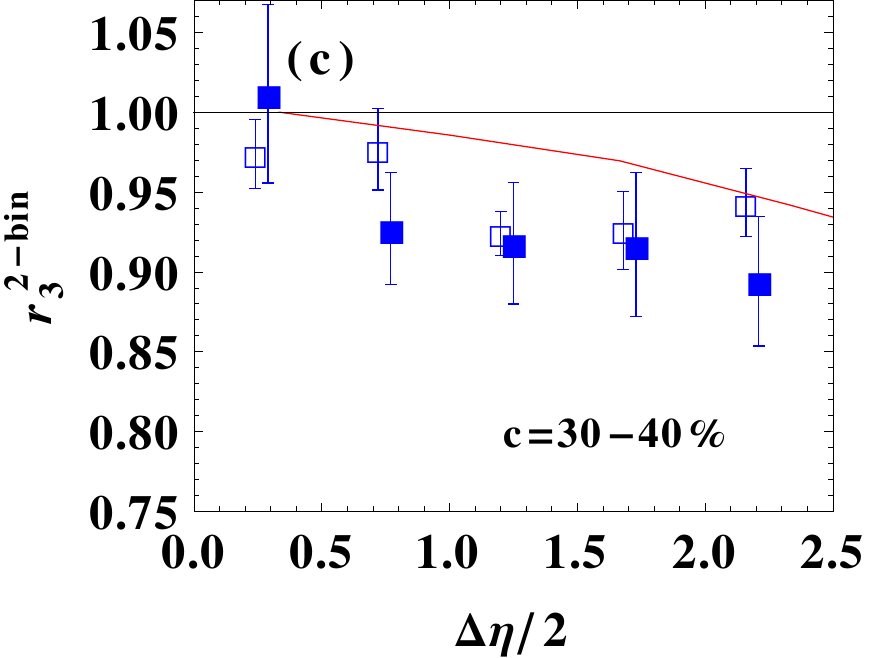} \hspace{7mm}
\includegraphics[width=0.35 \textwidth]{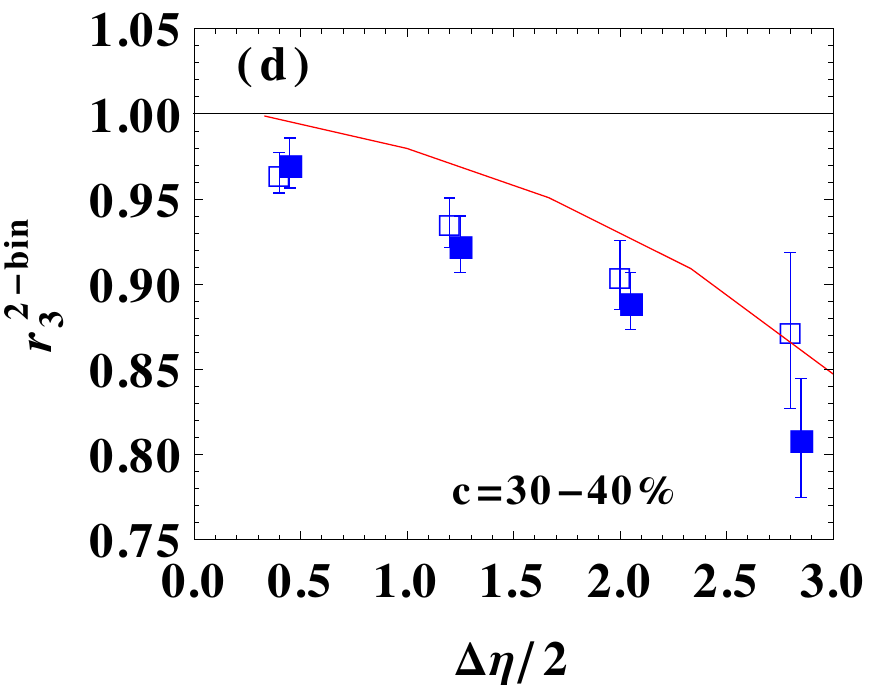}
\caption{(color online) Same as Fig.~\ref{fig:tv2} but for the triangular flow torque measure  $r_3^{\rm 2-bin}$.
\label{fig:tv3}}
\end{center}
\end{figure*}

Our numerical simulations are carried out for the same bin definition and kinematic cuts similar to Ref.~\cite{Khachatryan:2015oea}:
$-2.4 < \eta^a < 2.4$, $4 < \eta^b <5$, $0.3 < p_T^a < 3$~GeV, $0 < p_T^b < 3$~GeV.
The results are shown in Fig.~\ref{fig:3bin} for two sample centralities (0-5\% and 30-40\%) and for $n=2$ (elliptic flow) and $n=3$ (triangular flow). 
First, we note that while the statistical errors are large due to 
limited statistics, one may clearly see a qualitative agreement 
of the model points with the 
experiment. Also, the no-torque simulation (empty squares) is compatible with 1 within the statistical uncertainties, in agreement with the discussion above. 

To circumvent the limits 
of the finite statistics, we have also carried out the analysis with {\em cumulative} events (solid lines in Fig.~\ref{fig:3bin}). In this case 
we evaluate $v_{n\Delta}$ by summing up hadrons from all {\tt THERMINATOR} events simulated on top of a given hydro event. That way we eliminate the 
fluctuations coming from a 
finite number of the produced particles and minimize the non-flow effects. This trick is possible in model calculations, 
as opposed to the experiment, since in each {\tt THERMINATOR} event run on top of a given 
hydrodynamic event we have a fixed geometry of the collision.  We note that the result of the analysis with the cumulative events 
yields a smaller torque than the experiment. 

\section{2-bin correlations  \label{sec:2bin}}

Next, we pass to the results for the torque measure introduced in Ref.~\cite{Bozek:2010vz}.
It is defined via two-particle cumulants with particles belonging to bins $F$ and $B$:
\begin{eqnarray}
\!\!\!\!\!\!\!\!\!\! r_2^{\rm 2-bin} &\equiv& \frac{\langle \langle e^{i n(\phi_F-\phi_B)} \rangle \rangle}
{\sqrt{ \langle \langle e^{i n(\phi_{F,1}-\phi_{F,2})} \rangle \rangle \langle \langle e^{i n(\phi_{B,1}-\phi_{B,2})} \rangle \rangle} }. \label{eq:c2}
\end{eqnarray}
This measure is sensitive to the non-flow effects as well, 
hence is useful to estimate  such contributions to the torque effect.

The results of our simulations are shown in Figs.~\ref{fig:tv2} and \ref{fig:tv3}  for the cases of the elliptic and triangular flow, respectively. 
The cuts are $0.15 < p_T < 3$ ~GeV for RHIC and $0.3 < p_T < 3$ ~GeV for the LHC.
The corresponding calculation with cumulative events is indicated with the solid line. We note that due to non-flow effects the results for the case without the torque (empty 
squares) are significantly below unity. The presence of torque (filled squares) lowers further the results, but this effect would be 
difficult to assess experimentally. We note that the difference of the torque and no-torque curves is very close to the 
corresponding difference for the cumulative events (where the no-torque result overlaps with unity). We also note that the predicted effect of the torque is significantly higher at 
RHIC than at the LHC, and is stronger for $n=3$ than for $n=2$. 

Comparison of the results of Fig.~\ref{fig:3bin} with Figs.~\ref{fig:tv2} and \ref{fig:tv3} shows that the 3-bin measure is advantageous with regard to the 2-bin measures, as the non-flow 
effects are greatly reduced.

\section{Conclusions \label{sec:conclusions}}

The very accurate studies at the LHC make the flow correlation studies feasible and practical. Indeed, the error bars in the 
CMS analysis of Ref.~\cite{Khachatryan:2015oea} are much smaller than the size of the investigated effect. Our event-by-event simulations 
confirm that the interpretation of the behavior of the CMS measure $r_n(\eta^a, \eta^b)$ via the torque effect proposed in Ref.~\cite{Bozek:2010vz}
is natural: the decorrelation of the event-plane orientations between bins distant in pseudorapidity follows from fluctuations of the left- and right-going 
wounded nucleons and their asymmetric emission profiles.

Nevertheless, precise modeling of the torque effect may be a numerical challenge, as it requires a very large event-by-event statistics. 
Yet, with the number of events at our disposal, we can clearly see several features: the effect is larger at RHIC than at the LHC, 
it is also larger for the  triangular flow than for elliptic flow. The presence of non-flow affects the two-bin measures, while the application of the 
three-bin method proposed by the CMS~\cite{Khachatryan:2015oea} minimizes these effects.

%The effect of the subleading component or event-plane 
%decorrelation is visible for collisions at RHIC energies. For Pb-Pb collisions at the LHC, event-plane decorrelation can be measured 
%for the third order event plane at all centralities and for the second order event plane in central collisions, using an observable
%constructed with flow vectors from three interval in pseudorapidity. 

In future studies one could also apply to our hydrodynamic model the Principal Component Analysis (PCA), recently brought up in Refs.~\cite{Bhalerao:2014mua,Bhalerao:2014xra}. 
This would allow one to quantify the torque effect (and other correlation effects) 
from a more general perspective of the multivariate analysis, similar to the case of the AMPT model~\cite{Lin:2004en} investigated in  
Ref.~\cite{Bhalerao:2014mua,Bhalerao:2014xra}. In analogy to the two-bin measure $r_2^{\rm 2-bin}$, dominated by non-flow effects, we expect that these effects 
will also influence the PCA technique based on covariance matrix composed of two-bin observables. Thus it is important to compare model predictions with and without the torque 
effect in the fireball fluctuations, as done in the present work.

%However, as PCA is based on two-body correlations, 
%A recent proposition uses the principal component analysis to identify the dominant eigenmodes in the correlation 
%matrix, both for the  multiplicity and for the event-plane angles of second and third order.
%The PCA analysis of the correlations in pseudorapidity intervals reveals a weak subleading component for RHIC energies.

\begin{acknowledgments}

Research supported by the Polish Ministry of Science and Higher Education (MNiSW), by the National
Science Center grants DEC-2012/05/B/ST2/02528 and DEC-2012/06/A/ST2/00390, as well as by PL-Grid Infrastructure.

\end{acknowledgments}

\bibliography{hydr}

\end{document}